\documentclass[%
reprint,
 amsmath,amssymb,
 aps,
]{revtex4-2}

\usepackage{graphicx}
\usepackage{dcolumn}
\usepackage{bm}
\usepackage{hyperref}
\usepackage{amssymb}
\hypersetup{colorlinks=true,
    linkcolor=blue,
    citecolor=blue,
    filecolor=magenta,      
    urlcolor=blue,
    }
\usepackage{chemfig}
\usepackage{soul}
\usepackage[version=4]{mhchem}
\usepackage{subcaption}
\usepackage{graphicx}

\usepackage{float}
\NewDocumentCommand{\tens}{e{_^}}{%
  \mathbin{\mathop{\otimes}\displaylimits
    \IfValueT{#1}{_{#1}}
    \IfValueT{#2}{^{#2}}
  }%
}
\DeclareUnicodeCharacter{2212}{-}
\usepackage{pifont}
\newcommand{\cmark}{\ding{51}}%
\newcommand{\xmark}{\ding{55}}%

\begin{document}

\preprint{APS/123-QED}

\title{SHARC-VQE: Simplified Hamiltonian Approach with Refinement and Correction enabled Variational Quantum Eigensolver for Molecular Simulation 
\\
}
\author{Harshdeep Singh}
\email{harshdeeps@kgpian.iitkgp.ac.in}
\affiliation{%
 Center of Computational and Data Sciences, Indian Institute of Technology, Kharagpur, India
}%
\author{Sonjoy Majumder}%
\email{sonjoym@phy.iitkgp.ac.in}
\affiliation{%
  Department of Physics, Indian Institute of Technology, Kharagpur, India}%
\author{Sabyashachi Mishra}%
\email{mishra@chem.iitkgp.ac.in}
\affiliation{%
  Department of Chemistry, Indian Institute of Technology, Kharagpur, India}%

\date{\today}

\begin{abstract}
Quantum computing is finding increasingly more applications in quantum chemistry, particularly to simulate electronic structure and molecular properties of simple systems. The transformation of a molecular Hamiltonian from the fermionic space to the qubit space results in a series of Pauli strings. Calculating the energy then involves evaluating the expectation values of each of these strings, which presents a significant bottleneck for applying variational quantum eigensolvers (VQEs) in quantum chemistry. Unlike fermionic Hamiltonians, the terms in a qubit Hamiltonian are additive. This work leverages this property to introduce a novel method for extracting information from the partial qubit Hamiltonian, thereby enhancing the efficiency of VQEs. This work introduces the SHARC-VQE (Simplified Hamiltonian Approximation, Refinement, and Correction-VQE) method, where the full molecular Hamiltonian is partitioned into two parts based on the ease of quantum execution.  The easy-to-execute part constitutes the Partial Hamiltonian, and the remaining part, while more complex to execute, is generally less significant. The latter is approximated by a refined operator and added up as a correction into the partial Hamiltonian. SHARC-VQE significantly reduces computational costs for molecular simulations. The cost of a single energy measurement can be reduced from $\mathcal{O}(\frac{N^4}{\epsilon^2})$ to $\mathcal{O}(\frac{1}{\epsilon^2})$ for a system of $N$ qubits and accuracy $\epsilon$, while the overall cost of VQE can be reduced from $\mathcal{O}(\frac{N^7}{\epsilon^2})$ to $\mathcal{O}(\frac{N^3}{\epsilon^2})$. Furthermore, measurement outcomes using SHARC-VQE are less prone to errors induced by noise from quantum circuits, reducing the errors from 20-40\% to 5-10\% without any additional error correction or mitigation technique. Additionally, the SHARC-VQE is demonstrated as an initialization technique, where the simplified partial Hamiltonian is used to identify an optimal starting point for a complex problem. Overall, this method improves the efficiency of VQEs and enhances the accuracy and reliability of quantum simulations by mitigating noise and overcoming computational challenges.


\end{abstract}

\maketitle


\section{Introduction}
\label{sec:intro}
\setcounter{equation}{0}
Quantum computing utilizes the concepts of quantum mechanics to carry out computations that are fundamentally distinct from those performed by classical computing~\cite{qcintro1, qcintro2}. Quantum computing relies on qubits, which can concurrently exist in numerous states through superposition, unlike classical bits. In addition, qubits can become entangled, which implies that the state of one qubit can be influenced by the state of another qubit, regardless of the distance between them. The phenomenon of entanglement, along with superposition, allows quantum computers to simultaneously generate and manipulate a large quantity of data, resulting in potential exponential acceleration for specific problems~\cite{Advantage1, farhi2019quantum, jaramillo_2016_quantum, first}. Its capacity for catalyzing substantial change is notably evident in the discipline of chemistry~\cite{qintro1, qintro3, qintro6, cao_2019_quantum}. 

Although there have been significant improvements in the classical simulation of quantum chemical systems~\cite{intro1, intro2}, classical computing faces difficulties in accurately simulating the intrinsic quantum properties of electrons, particularly in large molecules and materials. As a result, its ability to predict and simulate molecular interactions with a high level of accuracy is limited~\cite{challenge2, challenge3}. Quantum computers are particularly suitable for simulating structure and properties in the areas of catalysis, materials research, and drug discovery~\cite{drug1, drug2, drug3}. In this early era of quantum computing, hybrid classical-quantum algorithms such as Variational Quantum Eigensovlers (VQE) have become the most efficient methods for doing molecular electronic structure calculations~\cite{vqa, qintro4, qintro5}. These algorithms have undergone experimental testing on several quantum hardware, such as photonic processors (photonic) and trapped-ion processors~\cite{trapion}. The use of efficient ansatzes such as the unitary coupled cluster methods~\cite{qintro2, qintro7}, and the advances in error-resistant algorithms, and quantum error correction~\cite{error1, error2, error3, halder_2023_development} add further to the utility of these variational algorithms. 
\par

Despite the potential for revolutionizing computing, the current generation of quantum computers, the so-called Noisy Intermediate-Scale Quantum (NISQ) computers, are susceptible to external noise and face many other challenges to be fully utilized. The major bottlenecks for scaling up the size and length of quantum computations are the availability of a limited number of qubits and the short coherence time of the employed qubits. Recently, we benchmarked the performance of different classical optimizers for quantum chemistry calculation with variational quantum algorithms~\cite{hd}. Despite using the best available classical optimizer, the error in the ground-state energy calculation across different simple molecules in the presence of noise was about 10-20$\%$, which does not bode well for the applicability and scalability of the VQAs. Two major hurdles that contribute the most to the noise are identified as the high-depth ansatz, which increases the number of quantum gates in the quantum circuits, and the size of the molecular Hamiltonian, which increases the number of measurements of the quantum circuits. A lot of work has been done in the modification and construction of efficient ansatz specific to quantum chemistry problems with VQAs~\cite{ansatz1, ansatz2, ansatz3, ansatz4, ansatz5}. In the pursuit of increasing the accuracy and flexibility of the wavefunctions, these ansatzes invariably increase the depth of the quantum circuits. This adds further to the cost and error in the measurements needed to estimate the expectation value of the Hamiltonian using VQE~\cite{resource1, resource2, resource3, resource4}.

Quantum error mitigation (QEM) can be one of the potential solutions for overcoming the measurement problem, and a significant amount of work has been done in that area~\cite{qem1, qem2, qem3, qem4, qem5}. However, most of the error mitigation techniques significantly increase the resource requirement of the VQEs. To overcome this, scalable QEM techniques have been devised~\cite{qem6, qem7}. There have also been efforts to quantify and mitigate the errors introduced by the quantum gates in the VQEs~\cite{qem8, qem9}.  However, the focus has to be placed on the growing size of the molecular Hamiltonian and the consequent exponential rise in resource requirements, as that would negate all other endeavors in error mitigation. While some work has been done in devising efficient methods of qubit Hamiltonian measurement, many approaches and techniques are yet to be explored. A typical $N-$qubit Hamiltonian consists of several Pauli operators $P_i$ with their respective coefficients $c_i$, such that, 
\begin{equation}
    H =\sum^{n}_{i=1}c_iP_i, \label{eq:sum_of_pdt_pauli}
\end{equation}
where each Pauli operator $P_i$ is a $N-$dimensional Pauli string. The number of terms in a $N$-qubit Hamiltonian scales as $\mathcal{O}(N^4)$. Each string is measured individually to determine its contribution to the overall energy, resulting in a high measurement overhead.  Since quantum measurements are probabilistic and noise-susceptible, they amplify inaccuracies when aggregating results from $\mathcal{O}(N^4)$ Pauli strings. 

In this work, we introduce a simplified Hamiltonian approach to improve the efficiency of VQE. The method employs a divide-and-conquer strategy, partitioning the full molecular Hamiltonian into smaller fragments~\cite{partialppr} based on their ease of computation and contribution to the overall Hamiltonian. While the simplified Hamiltonian containing significant and easy-to-compute fragments is solved exactly, the insignificant and difficult-to-compute fragments are approximated by some easy-to-compute refined operators within the VQE scheme. Post variational optimization, the final corrected energy is obtained by a subtractive scheme where the contribution of the refined operators is replaced by the true operators. This Simplified Hamiltonian Approach with Refinement and Correction enabled VQE (SHARC-VQE) is tested for ground state energy and wave functions of molecules with varying sizes, described by 4 to 10 qubits in ideal and noisy quantum environments. The approach is also extended to the excited states of molecules. The versatility of the method as an initialization technique is demonstrated for a general Fermi-Hubbard model. Further study also revealed that replacing a full Hamiltonian with a simpler partial Hamiltonian does not worsen the performance of the VQEs against the barren plateau problems, as suspected in literature~\cite{bpmain}.

\section{SHARC-VQE}
\subsection{Algorithm}
The Hamiltonian of a molecular system with $N$ electrons and $M$ nuclei under the Born-Oppenheimer approximation is given by, 
\begin{equation}
H=-\sum_{i=1}^{N} \left( \frac{\nabla_i^2}{2}  + \sum_{A=1}^{M} \frac{Z_A}{r_{iA}} \right) + \sum_{j>i}\frac{1}{r_{ij}} + \sum_{B>A}\frac{Z_AZ_B}{R_{AB}},   
\label{eq:3}
\end{equation}
where the first sum represents the one-electron operators (electron kinetic energy and electron potential energy), the second sum accounts for the two-electron interaction operators, and the last sum reflects the constant inter-nuclear repulsion. The molecular Hamiltonian can be expressed equivalently in the Fock space in the second quantization form as,
\begin{equation}
    H = \sum_{p,q} h_{pq} a_p^\dagger a_q + \frac{1}{2}\sum_{p,q,r,s}h_{pqrs}a_p^\dagger a_q^\dagger a_s a_r.
    \label{eq:8}
\end{equation}
In a given electronic basis set $\{X(\Vec{x})\}$, the one- and the two-electron integrals ($h_{pq}$ and $h_{pqrs}$, respectively) are given by, 
\begin{equation}
    \displaystyle h_{pq} = \int d \vec{x} X_p^*(\vec{x})\left(-\frac{\nabla^2}{2}-\sum_{A}\frac{Z_{A}}{r_{A \vec x}}\right)X_q(\vec{x})
    \label{eq:9}
\end{equation}
and 
\begin{equation}
    h_{pqrs} = \iint d \vec{x_1}d \vec{x_2}\frac {X_p^*(\vec{x_1})X_q^*(\vec{x_2})X_r(\vec{x_1})X_s(\vec{x_2})}{r_{12}}.
\end{equation}
The Fock space representation of the wavefunction is expressed in terms of occupation numbers vector,
\begin{equation}
    \left| s \right \rangle = \left| s_1, \cdots, s_p, \cdots ,s_N \right \rangle, 
    \label{eq:4}
\end{equation}
with $s_p= \{0, 1\}$ represents if the $p^{th}$ spin-orbital is unoccupied or occupied, respectively. The fermionic operators $a^\dagger_p$ and $a_p$ in the Fock-space representation are then employed to create and annihilate an electron in the $s_p$ spin-orbital by,
\begin{eqnarray}
    a^\dagger_{p} | s\rangle &=& (1-\delta_{s_p,1}) \Gamma^s_p \mid s_1, \cdots ,1_p, \cdots, s_n\rangle \label{eq:6} \\
    a_p |  s\rangle &=& \delta_{s_p,1} \Gamma^s_p | s_1, \cdots ,0_p,\cdots ,s_n\rangle. 
    \label{eq:7}
\end{eqnarray}
Here $\displaystyle{\Gamma^s_p = (-1)^{ \sum_{m<p}s_m}}$ ensures the antisymmetric nature of the wave function with respect to electron exchange and the delta function $\delta_{s_p,1}$ enforces the Pauli's exclusion principle. 

In the Fock-space representation, the one-to-one correspondence between the fermionic state $|s\rangle$ and the qubit state $|q\rangle$ is straightforward, i.e.,
\begin{equation}
    | s\rangle =  | s_1, \cdots, s_N\rangle \Longleftrightarrow	 |  q\rangle = | q_1, \cdots ,q_N\rangle,
    \label{eq:5}
\end{equation}
where, the occupancy of a spin-orbital $s_p$ ($\{0,1\}$) maps to the binary state ($\{\uparrow , \downarrow\}$) of qubit $q_p$. The transformation between the fermionic space and the qubit space can be achieved by various transformation schemes, such as the Jordan-Wigner~\cite{jw}, Parity~\cite{parity}, and Brayvi-Kitaev~\cite{bk} schemes. In the Jordan-Wigner transformation, the fermionic operators are expressed in terms of the  Pauli spin operators ($X, Y, Z$) as, 
\begin{eqnarray}
    a_p^\dagger  &=& \frac{1}{2} (X_p - iY_p) \otimes_{q<p} Z_q \nonumber \\
    a_p &=& \frac{1}{2} (X_p + iY_p) \otimes_{q<p} Z_q \label{eq:tr_rule}.
\end{eqnarray}
Using the above expressions in Equation~\ref{eq:8}, the molecular Hamiltonian in qubit representation appears as a sum-of-products of Pauli operators (Equation~\ref{eq:sum_of_pdt_pauli}). For example, the Hamiltonian for H$_2$ in a 4-qubit space appears as~\cite{hydrogen}, 
\begin{equation}
    H_{\rm qubit}^{\rm H_2} = -0.81054 \ IIII + \cdots + 0.12091 \  ZZII,
\end{equation}
with the complete list given in Table~\ref{tab:h2full}. A quantum measurement of the above Hamiltonian involves the measurement of each Pauli string separately (i.e., 15 separate measurements in this case).

\begin{table}[htbp]
\caption{\label{tab:h2full}Hamiltonian of H$_2$ in 4-qubit space with a general form ($H=\sum_i c_iP_i$, where $P_i$s are the 4-qubit gates with coefficient $c_i$ (in au). The qubit Hamiltonians of the other systems considered in this work are given in the supporting information.}
\begin{center}
\begin{tabular}{ccc|ccc}
\hline
\hline
Index$^{\mathrm{a}}$  & Coefficient  & Gates & Index$^{\mathrm{a}}$  & Coefficient  & Gates\\
\hline
\multicolumn{6}{c}{Full Hamiltonian of H$_2$} \\ \hline 
1    &  $-$0.81054  & IIII  & 9     & $-$ 0.04523 & IXZX\\
2    & $+$0.17218   & IIIZ  &  10   & $-$ 0.04523& ZXZX\\
3    & $-$0.22575   & IIZZ  &  11   & $+$ 0.04523 & IXIX\\
4    & $+$ 0.17218   & IZZI  &  12   & $+$ 0.16614 & ZZIZ\\
5    & $-$ 0.22575   & ZZII  &  13   & $+$ 0.16614 & IZIZ\\
6    & $+$ 0.12091   & IIZI  &   14  & $+$ 0.17464 & ZZZZ\\
7    & $+$ 0.16892   & IZZZ  &  15   & $+$ 0.12091 & ZIZI\\
8    & $+$ 0.04523  & ZXIX  & & & \\
\hline
\multicolumn{6}{c}{Only one-electron operators of H$_2$} \\ \hline 
1    &  $-$ 1.72823 & IIII &4    & $+$ 0.62816   & IZZI \\
2    & $+$ 0.62816   & IIIZ &5    & $+$ 0.23594   & ZZII \\
3    & $+$ 23594   & IIZZ &&&\\ 
\hline
\multicolumn{6}{c}{Only two-electron operators of H$_2$} \\ \hline 
1    &  $+$0.91768  & IIII  & 9     & $-$ 0.04523 & IXZX\\
2    & $-$0.45598   & IIIZ  &  10   & $-$ 0.04523& ZXZX\\
3    & $-$0.46170   & IIZZ  &  11   & $+$ 0.04523 & IXIX\\
4    & $-$ 0.45598   & IZZI  &  12   & $+$ 0.16614 & ZZIZ\\
5    & $-$ 0.46170   & ZZII  &  13   & $+$ 0.16614 & IZIZ\\
6    & $+$ 0.12091   & IIZI  &   14  & $+$ 0.17464 & ZZZZ\\
7    & $+$ 0.16892   & IZZZ  &  15   & $+$ 0.12091 & ZIZI\\
8    & $+$ 0.04523  & ZXIX  & & & \\ \hline 
\multicolumn{6}{l}{$^{\mathrm{a}}$The sequence follows the Qiskit ordering.}
\end{tabular}
\end{center}
\end{table}

\begin{figure}[!]
 \includegraphics[width=0.4\textwidth]{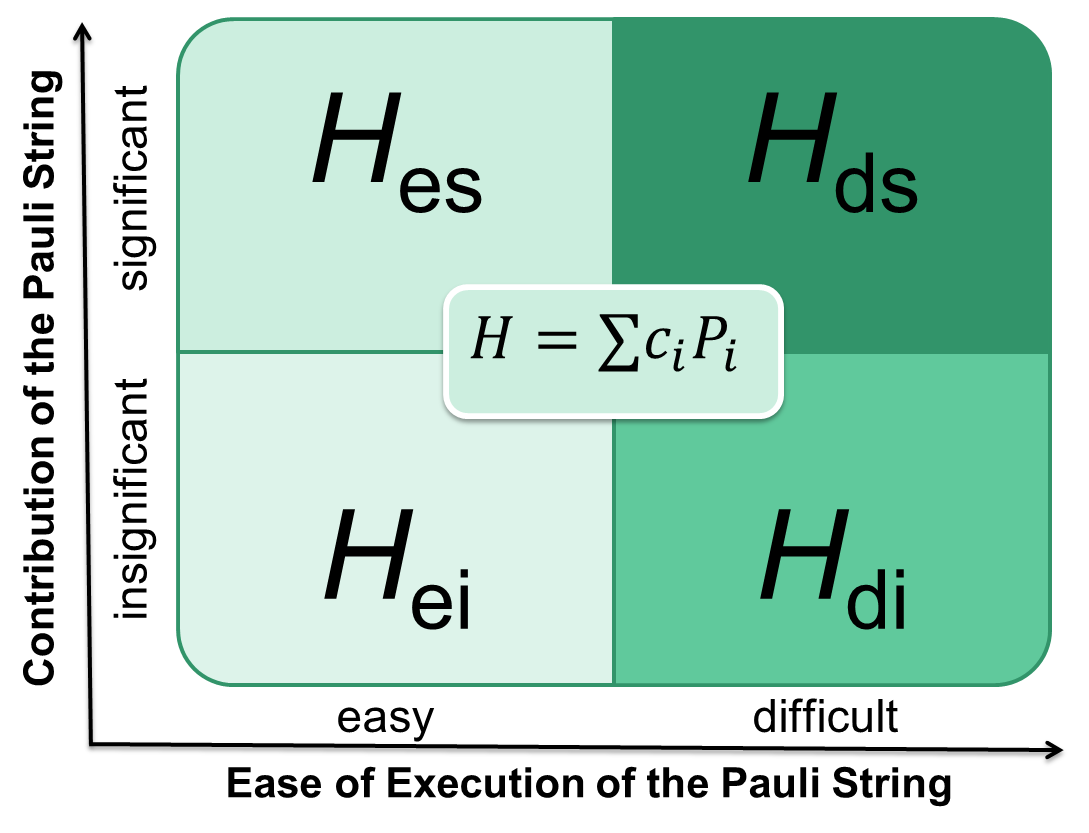}
 \caption{\label{fig:divisions} Separation of the qubit Hamiltonian based on the ease of execution of the Pauli strings and their significance indicated by their respective coefficients.}
\end{figure}

The most conspicuous way to divide the molecular Hamiltonian would be by transforming the one and two-electron terms in the fermionic Hamiltonian (Equation~\ref{eq:8}) separately via the Equation~\ref{eq:tr_rule} (see Table~\ref{tab:h2full}). It is obvious that ignoring a particular type of interaction altogether leads to highly incorrect results (see Figure~S1 in the supporting information). 
Hence, instead of ad-hoc fragmenting the fermionic Hamiltonian, we explore a systematic approach to selecting terms from the qubit Hamiltonian. The sum-of-products form of the qubit Hamiltonian (Table~\ref{tab:h2full}) can be fragmented based on two factors, namely, the contribution of each term to the overall Hamiltonian ($c_i$) and the cost of execution of the Pauli string as described in FIG.~\ref{fig:divisions}. 
Considering a certain cut-off value $c_0$, one can divide the terms in the Hamiltonian as significant ($c_i \ge c_0$) and insignificant ($c_i < c_0$). The Pauli strings with exclusively I and Z gates are easy to execute since all such operators commute with each other and require no additional basis transformation for measurement. Therefore, the entire group can be measured using a single quantum circuit. On the other hand, each term containing X or Y gate(s) requires appropriate basis transformation, thus raising the execution cost. Therefore, the qubit Hamiltonian of any system can be divided into four segments, e.g., the significant and easy-to-execute terms ($H_{\rm es}$), the significant and hard-to-execute terms ($H_{\rm ds}$), the insignificant and easy-to-execute terms ($H_{\rm ei}$), and the insignificant and hard-to-execute terms ($H_{\rm di}$), see FIG.~\ref{fig:divisions}. 

The qubit Hamiltonian can be simplified in such a way that requires the minimum number of circuit executions, thereby reducing the quantum load of the VQA. The simplest way of achieving this is to consider a partial Hamiltonian $H_{\rm p}$ (as $H_{\rm es}+H_{\rm ei}$), by including the terms with I or Z gates alone, notwithstanding their relative contribution. After all, a single circuit measurement shall suffice their evaluation.  A VQE optimization using the partial Hamiltonian yields the wave function in terms of the optimized rotation parameters $\{\theta^p\}$ and energy $E^\prime$, such that,  
\begin{equation}
H_{\rm p} | \Psi_{\rm p} (\theta^p) \rangle =E^\prime  |\Psi_{\rm p} (\theta^p) \rangle.
\end{equation}
The reliability of the state of the system optimized with the partial Hamiltonian needs to be tested against the results from VQE optimization with the full Hamiltonian within a desired level of accuracy ($\epsilon$). In other words, we ask if, 
\begin{equation}
\left| \langle \Psi_{\rm full} \mid H_{\rm full} \mid \Psi_{\rm full} \rangle  - \langle  \Psi_{\rm partial} \mid H_{\rm full} \mid \Psi_{\rm partial} \rangle \right | \le \epsilon 
\end{equation}
In most cases, the above relation is unlikely to hold true as the $H_{\rm p}$ has left out the $H_{\rm di}$ and $H_{\rm ds}$ terms of the Hamiltonian. The terms that are left out mainly arise from the two-electron interactions. Their number is expected to grow significantly with molecular size. On the other hand, the inclusion of the difficult-to-execute terms $H_{\rm d}$ within the VQE scheme offers a marginal contribution to the overall energy (note the small coefficients of the $H_{\rm d}$ terms in Table~\ref{tab:h2full}) all the while introducing a significant amount of measurement noise. Hence, it is not only necessary to add corrections to the energy obtained from partial Hamiltonian optimization but also to devise a method to incorporate the information of the left-out terms in some way in the VQE simulation itself. 

One way of achieving this is to approximate the difficult-to-execute part of the Hamiltonian with an easy-to-execute operator ($H^\prime_d$). Finding $H^\prime_d$ that can approximate $H_d$ at all possible points can be tedious. If $|\Psi(\theta^F)\rangle$ is the final optimized state at the end of a VQE simulation with the full Hamiltonian, we need the operator $H^\prime_d$ such that,
\begin{equation}
    \langle \Psi(\theta^F) | H_d |\Psi(\theta^F) \rangle \approx \langle \Psi(\theta^F) | H^{'}_d |\Psi(\theta^F) \rangle.
\end{equation}
Of course, finding $\theta^F$ is also an expensive task by itself. Under the assumption that the true state of a molecular system is not far from the Hartree-Fock state $|\Psi(\theta^{\rm HF})\rangle$, we can construct a refined operator as a sum of one or more easy-to-evaluate Pauli gates whose coefficients can be fitted to reproduce the expectation value of the $H_{\rm d}$ operator in the close vicinity of the Hartree-Fock state, that is, 
\begin{multline}
       \langle \Psi(\theta^{\rm HF}+\delta \theta) | H_d |\Psi(\theta^{\rm HF}+\delta \theta) \rangle \approx  \\ \langle \Psi(\theta^{\rm HF}+\delta \theta)) | H^\prime_d |\Psi(\theta^{\rm HF}+\delta \theta) \rangle 
\end{multline}
where, $\delta \theta$ is a small random perturbation. The advantage of invoking the HF state lies in the fact that it is very easy to initialize the HF state in a quantum circuit, 
Based on these assumptions, we introduce the refined operator $H^{\prime \ ij}_d$, where $i$ represents the number of easy-to-execute Pauli strings employed in place of the $H_{\rm d}$ terms and $j$ indicates the number of points in the close vicinity of HF state at which the operator has been approximated. Using the refined operator $H^{\prime \ ij}_d$, the refined partial Hamiltonian can be written as,
\begin{equation}
    H_p^{ij} = H_p + H^{\prime \ ij}_d
\end{equation}
The VQE optimization with the refined Hamiltonian  $H_p^{ij}$ yields the final optimized parameters $\theta^p$. The final energy is then estimated from a subtractive correction scheme in which the contribution of the refined operator  $H^{\prime \ ij}_d$ to the overall energy is substituted by the energy contribution of the true operator  $H_{\rm d}$, i.e.,

\begin{equation}
    E^{ij}_{\rm SHARC-VQE} = \left \langle \Psi(\theta^{p}) \left| H_p^{ij} -  H_d^{\prime \ ij}  +  H_d \right|\Psi(\theta^{p}) \right \rangle. \label{eq:en_sharc-vqe} 
\end{equation}

Equation~\ref{eq:en_sharc-vqe} ensures that the final energy accounts for the contribution from the true operator $H_{\rm d}$, whereas, during the iterative procedure of the VQE optimization, its approximate form ($H_d^{'ij}$) is used for state optimization. This simplified Hamiltonian approach (SHARC-VQE) can be employed for any partial Hamiltonian by replacing the left-out terms with some easy-to-compute refined operators during VQE optimization, and by correcting the final energy according to Equation~\ref{eq:en_sharc-vqe}. In a similar way, the $H_p^{00}$ method can also be defined, where the approximation operator is not used, but the contribution from $H^{d}$ is added at the end. The schematic of the SHARC-VQE algorithm is presented in FIG.~\ref{fig:gs_ideal}.  

\begin{figure*}[!]
 \includegraphics[width=\textwidth]{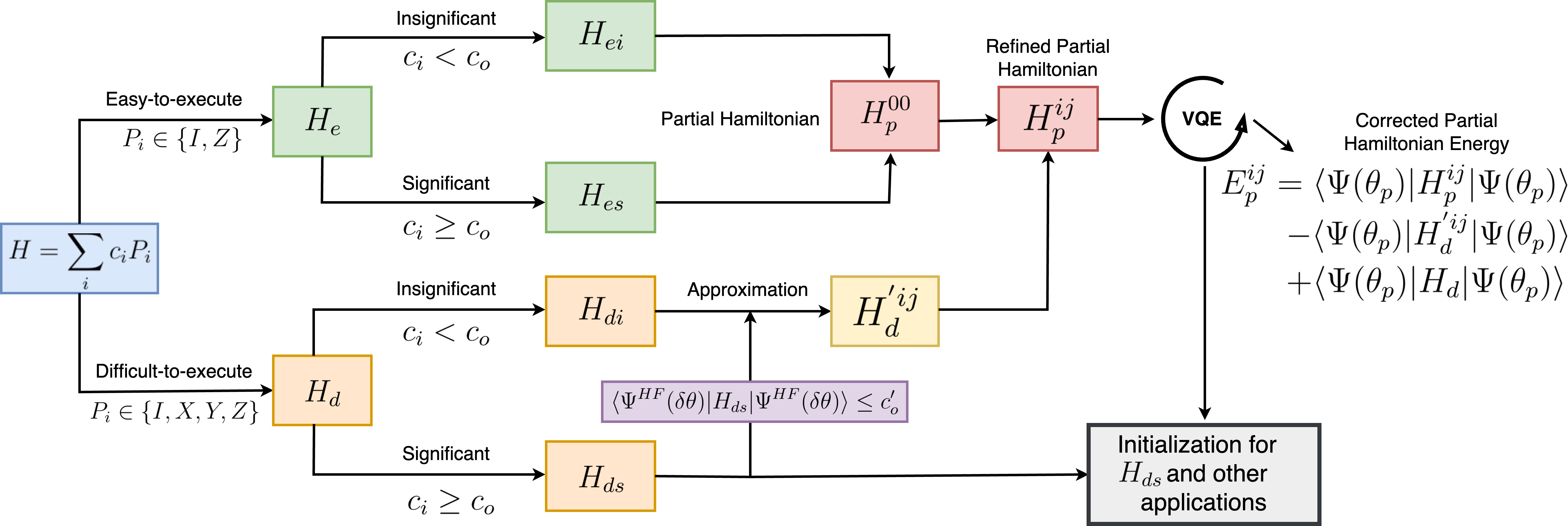}
 \caption{\label{fig:workflow}Workflow diagram for SHARC-VQE.}
\end{figure*}

\subsection{Implementation}


\begin{table*}
\caption{\label{tab:correction} The number of terms in different divisions of the Hamiltonian (FIG.~\ref{fig:divisions}) in different molecules. The terms with coefficients less than 0.01 au are treated as insignificant. The refined operator  ($H_d^{\prime 11}$ and  $H_d^{\prime 12}$) used in different molecules are given with coefficient and Pauli string.}
\begin{ruledtabular}
    \begin{tabular}{cccccccc}
Molecule & \# & \multicolumn{4}{c}{\# Terms in}  & $H_d^{\prime 11}$ & $H_d^{\prime  12}$ \\ \cline{3-6}
         & Qubits & $H_{\rm es}$  & $H_{\rm ei}$ & $H_{\rm ds}$ & $H_{\rm di}$ && \\
\hline  
       H$_2$ & 4  & 11 & 0 & 4 & 0 &  $-0.03594$ ZIZI & $-0.05320$ ZIZI \\
       LiH & 6  & 22 & 0 & 24 & 72 & $-0.00208$ ZZIZZI & $-0.00253$ ZZIZZI \\
       BeH$_2$ & 8 & 37  & 0 & 16 & 52 & $-0.00300$ ZZZIZZZI & $-0.00454$ ZZZIZZZI \\
       H$_2$O & 10 & 56  & 0 & 60 & 136 & $-0.00646$ ZZZIIZZZII & $-0.00814$ ZZZIIZZZII \\ 
    \end{tabular}
\end{ruledtabular}
\end{table*}

All VQE simulations employed the standard unitary coupled cluster with singles and doubles (UCCSD) ansatz~\cite{ucc} with ideal and noisy quantum circuit settings. For noisy simulation, the quantum circuit simulator (qasm\_simulator) with additional noise embedded from a noise model sampled from the IBM Cairo quantum device is employed, which gives an equivalent performance of a realistic fault-tolerant quantum computer. Based on our earlier benchmarking study~\cite{hd}, we used the SLSQP optimizer~\cite{slsqp} for ideal simulations, and the SPSA optimizer~\cite{spsa} for the noisy simulations. In the ideal conditions, the maximum number of iterations for SLSQP was set to 200, while for SPSA in noisy conditions, it was set to 500~\cite{hd}.

We considered H$_2$, LiH, BeH$_2$, and H$_2$O with 4, 6, 8, and 10 qubits, respectively, to test the performance of SHARC-VQE. With the increase in the size of the molecule (and qubit), the total number of terms in the qubit Hamiltonian increases.  Consequently, the number of gates with high-cost execution increases rapidly (Table~\ref{tab:correction}). These terms are approximated by some easy-to-execute refined operators listed in Table~\ref{tab:correction}. The refined operators can be written as, 
\begin{equation}
    H^{\prime \ ij}_{d} = \sum^{i}_{k=1} c^{(j)}_k P_k  \label{eq:refined_op}
\end{equation}
where, the coefficients $c_k$ are evaluated by comparing the expectation values of $H^{\prime \ ij}_{d}$, and $H_{d}$ in the vicinity of the Hartree-Fock state, averaged at $j$ points. $P_k$ are chosen in accordance with the Hartree-Fock state of the molecule under consideration. For example, for the H$_2$ molecule, a single Pauli string is chosen as ZIZI, and the coefficient for $H^{\prime \ 11}_{d}$ is evaluated as,
\begin{equation}
    c^{(1)}_{1} =  \frac{\langle \Psi(\theta^{HF}+\delta \theta)) | H_d |\Psi(\theta^{HF}+\delta \theta) \rangle}{\langle \Psi(\theta^{HF}+\delta \theta) | ZIZI |\Psi(\theta^{HF}+\delta \theta) \rangle}  . \label{eq:eq18}
\end{equation}
Similarly, for $H^{\prime \ 12}_{d}$, the coefficient $c_{12}$ takes the average of the two values evaluated at two points $\theta^{HF}+\delta \theta$, and $\theta^{HF}+2\delta \theta$. $\delta \theta$ can be a randomly chosen small value (0.1 in the present study). It is straightforward to extend this approach to greater values of $j$ (i.e., averaged over the $j$ points) and of $i$ (i.e., number of distinct easy-to-evaluate refined Pauli operators in Equation~\ref{eq:refined_op}). In the present case, $i=1$ and $j=1$, 2 were found to be sufficient for the chosen problems. 



The performance of SHARC-VQE under ideal and noisy conditions is tested with respect to the total energy and wavefunction of the molecule obtained from VQE optimization with the full Hamiltonian. The accuracy of energy evaluation can be found by defining the relative energy error of the different methods,
\begin{equation}
    \Delta E = \bigg| \frac{E^{\rm VQE} - E^{\rm ref}}{E^{\rm ref}} \bigg|
    \label{eq:rel_err}
\end{equation}
where $E^{\rm VQE}$ is the energy evaluated by the VQE method, while  $E^{\rm ref}$ is the reference energy, generally evaluated exactly by the Numpy Eigensolver. In this case, the chemical accuracy can be defined as a dimensionless quantity,
\begin{equation}
    \rm{Chemical\; Accuracy} = \bigg| \frac{1.6 \, \rm{mH}}{E^{\rm ref}} \bigg|
    \label{eq:chemacc}
\end{equation}
To evaluate the accuracy of the wavefunction, we computed the fidelity of the approximate wavefunction ($| \Psi_{\rm approx.} \rangle$) with the exact wavefunction ($| \Psi_{\rm exact} \rangle$) by, 
\begin{equation}
    {\cal F} (| \Psi_{\rm approx.} \rangle, | \Psi_{\rm exact} \rangle) = \left| \langle \Psi_{\rm approx.}  | \Psi_{\rm exact}  \rangle \right|^2.
\end{equation}
The fidelity of two quantum states measures their degree of similarity with $0\le {\cal F} \le 1$. Any deviation from 1 indicates a loss of accuracy. 

\section{Results and Discussion}

\begin{figure*}[!]
 \includegraphics[width=\textwidth]{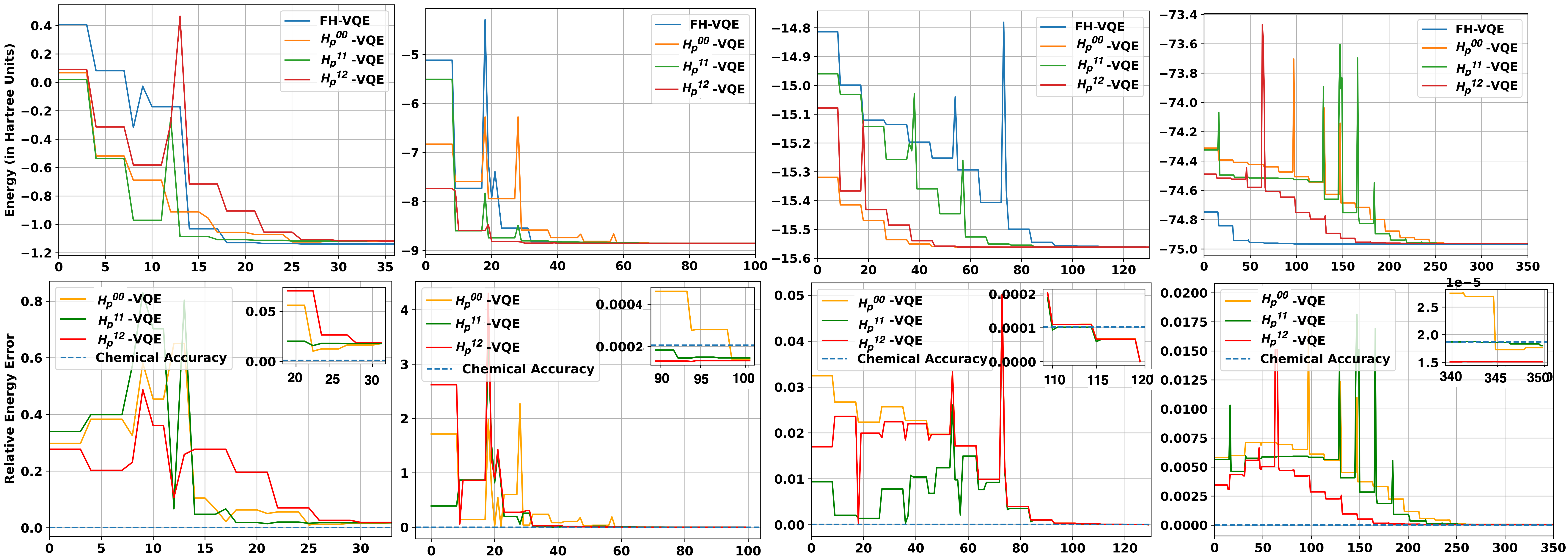} \\ 
  \includegraphics[width=\textwidth]{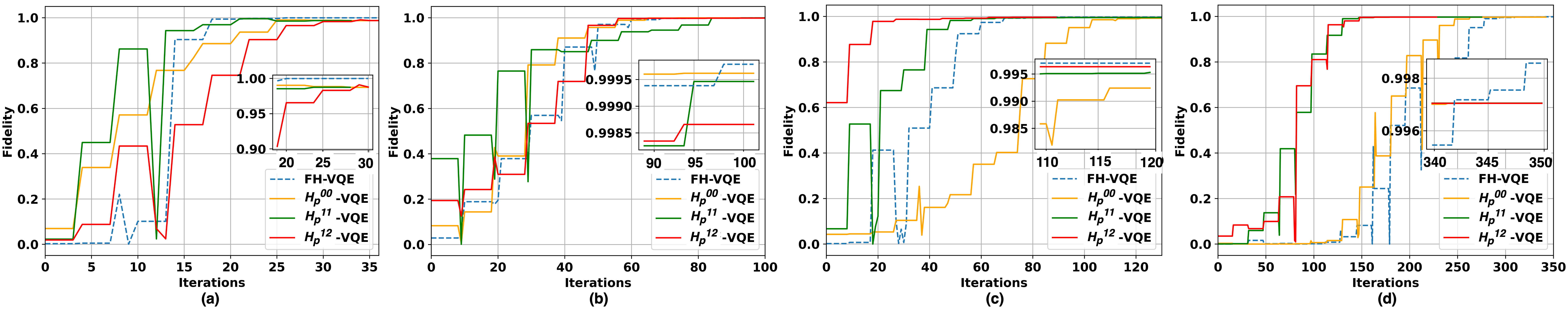}
 \caption{The ground state energy (upper row), the relative energy with respect to the corresponding converged ground state energy of the full Hamiltonian (middle row) and the fidelity of the wave function with respect to the wave function of the full Hamiltonian (bottom row) for (a) H$_2$, (b) LiH, (c) BeH$_2$, and (d) H$_2$O under ideal conditions. Chemical accuracy refers to Equation~\ref{eq:chemacc}. The inset shows a magnified view of the selected region.}\label{fig:gs_ideal}
\end{figure*}

\subsection{Performance of SHARC-VQE under Ideal Condition}

FIG.~\ref{fig:gs_ideal} compares the performance of all three variants of SHARC-VQE (differing by the refined operators employed to approximate the hard-to-execute terms of the Hamiltonian, Table~\ref{tab:correction}) under ideal conditions. In all cases, energy and the wave function converge within the chemical accuracy to their corresponding values with the full Hamiltonian VQE (FH-VQE). The relative energy errors show that 
different variants of SHARC-VQE require different numbers of iterations to achieve convergence. 
The corrected partial Hamiltonians, $H^{11}_p$, and $H^{12}_p$ do not show any particular difference from $H^{00}_p$ (without any approximation operator for $H_d$) in H$_2$. However, in LiH, BeH$_2$, the corrected partial Hamiltonians converge much faster than the uncorrected partial Hamiltonian and the full Hamiltonian (FIG.~\ref{fig:gs_ideal}). The approximation operators ensure that the SHARC-VQE method remains scalable to larger molecules. 

\subsubsection{Applicability of the Approximation Operators}
\begin{figure}[!]
 \includegraphics[width=0.5\textwidth]{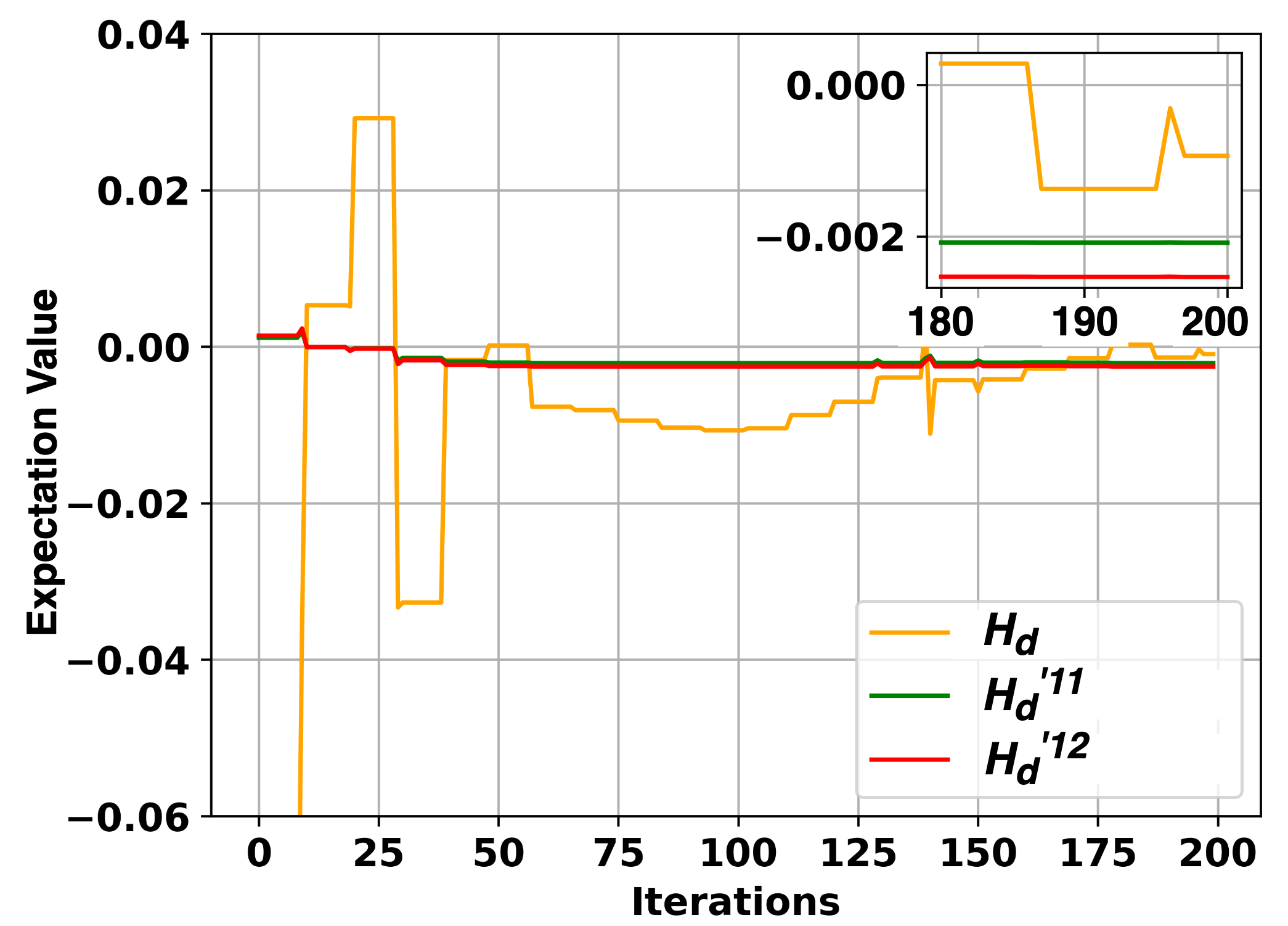}
 \caption{\label{fig:correc}The expectation values of the operators $H_{d}$, and $H^{'ij}_{d}$ along the iterations of a VQE simulation of the full Hamiltonian of LiH.}
\end{figure}

FIG.~\ref{fig:correc} shows the expectation values of the hard-to-execute terms in the qubit Hamiltonian ($H_{d}$) and those of their approximate form ($H^{'ij}_{d}$) along the VQE simulation of the full Hamiltonian of LiH. 
The error introduced by the approximation operator, $\epsilon({H^{'ij}_{d}})$, can be defined as the deviation of its expectation value from that of the true operator at a certain point $f$,
\begin{equation}
    \epsilon({H^{'ij}_{d}}) = |\langle H_{d} \rangle_f - \langle H^{'ij}_{d} \rangle_f|.
\end{equation}
It is found that the error is typically one or just a few orders less than the expectation value of the operator $H_{d}$, that is,  
\begin{equation}
    \mathcal{O}(\epsilon({H^{'ij}_{d}})) = \mathcal{O}(\langle H_{d} \rangle_f)\times 10^{-k}
\end{equation}
with, $k \in \{1,2\}$.

The hard-to-execute terms in the qubit Hamiltonian ($H_{\rm d}$) include Pauli strings with significant and insignificant coefficients. The insignificant part of this operator $H_{\rm di}$ can be safely estimated with the approximation operator.  However, the same may not hold up for the hard-to-execute terms with significant contribution ($H_{\rm ds}$). 
The actual contribution of a particular term in the Hamiltonian to the final energy depends on the expectation value of the corresponding Pauli string ($|\langle P_i \rangle| \le 1$), weighted by the coefficient ($c_i$). As long as the overall expectation value $|c_i\langle P_i \rangle|$ is less than a particular threshold value ($c'_0$, e.g., 1 mH as the chemical accuracy), the operator $H_{\rm ds}$ can be estimated using the approximation operator $H^{'ij}_d$. If that is not the case, one can use the SHARC-VQE method as an initialization technique to resolve the Hamiltonian (see Section~\ref{sec:initialization}).

\subsubsection{SHARC-VQE for Excited States with Variational Quantum Deflation}
While VQE offers the solution for the ground state of a problem, the solution of the excited states can be obtained with some modification to the VQE algorithm. One such algorithm employed in the current work is the variational quantum deflation (VQD) method~\cite{vqd}. 
Starting from the optimized VQE solution, the VQD method finds the $k^{th}$-excited state of the system by optimizing the parameters $\theta_k$ for the state $|\Psi_k\rangle$ such that the function $F(\Vec{\theta}_k)$ is variationally optimized, where
\begin{eqnarray}
      F(\Vec{\theta}_k) &=& \langle \Psi(\Vec{\theta}_k)|H|\Psi(\Vec{\theta}_k)\rangle + \sum_{j=0}^{k-1} \gamma_j |\langle \Psi(\Vec{\theta}_k)|\Psi(\Vec{\theta}_j) \rangle|^2
\nonumber  \\ 
        &=&
        E(\Vec{\theta}_k) + \sum_{j=0}^{k-1} \gamma_j |\langle \Psi(\Vec{\theta}_k)|\Psi(\Vec{\theta}_j) \rangle|^2.
        \label{eq:2}         
\end{eqnarray}
Here, $\{\Vec{\theta}_k\}$ is the optimized parameters of the $k^{th}$ energy state. VQD method optimizes $E(\Vec{\theta}_k)$ with an additional constraint that the current excited state $|\Psi(\theta_k)\rangle$ is orthogonal to the previous states $|\Psi(\Vec{\theta}_0)\rangle$, 
$\cdots$,  $|\Psi(\Vec{\theta}_{k-1})\rangle$. Here, $\gamma$ balances the contribution of each overlap term to the cost function and is generally computed as the mean square sum of the coefficients of the observable. This is equivalent to finding the ground state energy of a modified Hamiltonian at a stage $k$,
\begin{equation}
    H_k = H + \sum_{l=0}^{k-1} \gamma_l |l\rangle \langle l|
\label{eq:vqd_h}
\end{equation}
where, $|l\rangle$ is the $l^{th}$ eigenstate of the Hamiltonian $H$ with energy $E_l = \langle l|H|l\rangle $.
The VQD method has been successfully applied in various problems~\cite{huckel1, huckel2, huckel3}.
Here, we have extended the VQD approach within the SHARC scheme (FIG.~\ref{fig:workflow}), where the Hamiltonian  $H_k$ in Equation~\ref{eq:vqd_h} can be modified to,
\begin{equation}
    H^{ij}_{kp} = H^{ij}_p + \sum_{l=0}^{k-1} \gamma_l |l\rangle \langle l|
\label{eq:vqd_2}
\end{equation}
where, $H^{ij}_p$ is the partial Hamiltonian constructed from the initial full Hamiltonian $H$ of different molecules. 
No additional changes were made with regard to the $\sum_{l=0}^{k-1} \gamma_l |l\rangle \langle l|$ terms. In Table~\ref{tab:excited}, we demonstrate the performance of various SHARC-VQD schemes for the lowest four states of the systems under consideration. The results highlight that the SHARC-VQE can be an effective tool for exploring the entire energy spectrum of the molecules with the partial Hamiltonian $H^{ij}_{kp}$. The averaged relative energy energies with respect to the exact energies show that in all the cases, the VQD algorithm works equally well with the full and partial Hamiltonians. 


\begin{table*}
\caption{\label{tab:excited} The ground and excited state energy (in Hartree) and averaged relative energy error ([\ref{eq:rel_err}]) with full Hamiltonian VQE and SHARC-VQE $(H^{ij}_p)$ in an ideal quantum circuit simulator.}
\begin{ruledtabular}
    \begin{tabular}{c|cccccccc}

Molecule &  Method & $E_{gs}$ & $E_1$ & $E_2$ & $E_3$ & $E_4$ & $\Delta E$\\
 &  &  &  &  &  &  & $(\times  10^{-3})$\\
\hline  
\hline  
H$_2$ & Exact & $-$1.1373 & $-$0.5363 & $-$0.5363 & $-$0.5246 & $-$0.5246 & -\\
             & $H$ & $-$1.1373 & $-$0.5363 & $-$0.5291 & $-$0.5246 & $-$0.5125 & 3.0\\
               & $H_p^{00}$ & $-$1.1169 & $-$0.5340 & $-$0.5316 & $-$0.5287 & $-$0.5269 & 1.52\\
               & $H_p^{11}$ & $-$1.1170 & $-$0.5363 & $-$0.5293 & $-$0.5269 & $-$0.5124 & 1.08\\
               & $H_p^{12}$ & $-$1.1170 & $-$0.5363 & $-$0.5316 & $-$0.5291 & $-$0.5243 & 6.6\\
\hline  
LiH & Exact & $-$7.8636 & $-$7.7845 & $-$7.7845 & $-$7.7176 & $-$7.7176  & -\\
             & $H$ & $-$7.8634 & $-$7.7844 & $-$7.7443 & $-$7.7290 & $-$7.7174  & 1.2\\
               & $H_p^{00}$ &  $-$7.8633 & $-$7.7842 & $-$7.7551 & $-$7.7115 & $-$7.7176  & 0.6\\
               & $H_p^{11}$ &  $-$7.8341 & $-$7.7867 & $-$7.7528 & $-$7.7230 & $-$7.6848  & 1.6\\ 
               & $H_p^{12}$ &  $-$7.8633 & $-$7.7843 & $-$7.7675 & $-$7.7283 & $-$7.7148  & 0.6\\ 
\hline  
BeH$_2$ & Exact & $-$15.5643 & $-$15.3502 & $-$15.3502 & $-$15.2818 & $-$15.2818  & -\\
             & $H$ & $-$15.5625 & $-$15.3502 & $-$15.3213 & $-$15.3015 & $-$15.2736  & 0.4\\
               & $H_p^{00}$ &  $-$15.5612 & $-$15.3502 & $-$15.3317 & $-$15.2881 & $-$15.2285  & 0.8\\
               & $H_p^{11}$ &  $-$15.5612 & $-$15.3502 & $-$15.3228 & $-$15.2863 & $-$15.2876  & 0.2\\ 
               & $H_p^{12}$ &  $-$15.5612 & $-$15.3502 & $-$15.2864 & $-$15.2877 & $-$15.2849  & 0.8\\ 
\hline  
H$_2$O & Exact & $-$74.9702 & $-$74.6091 & $-$74.6091 & $-$74.5622 & $-$74.5622 & - \\	
             & $H$ &  $-$74.9661 & $-$74.5589 & $-$74.5530 & $-$74.4901 & $-$74.4826 & 0.2\\
               & $H_p^{00}$ & $-$74.9629 & $-$74.5673 & $-$74.5334 & $-$74.4605 & $-$74.4493 & 0.6\\
               & $H_p^{11}$ &  $-$74.9629 & $-$74.5290 & $-$74.5509 & $-$74.5337 & $-$74.5361 & 0.2\\ 
               & $H_p^{12}$ &  $-$74.9629 & $-$74.5199 & $-$74.5193 & $-$74.5055 & $-$74.4837 & 0.6 
\label{table4}
\end{tabular}
\end{ruledtabular}
\end{table*}

\subsection{Performance of SHARC-VQE under Noisy Condition}
Quantum noise is a fundamental hindrance in existing quantum technology, leading to inaccuracies in qubit states and operations, thus affecting the reliability of calculations. The noise arises from multiple sources, such as decoherence (which leads to the loss of the quantum state in qubits over time); gate errors (which occur during the implementation of quantum gates and result in imperfect operations); and measurement errors (which cause inaccuracies when determining the final state of qubits). The effect of quantum noise can be particularly damaging to the variational algorithms. The iterative process of VQAs, which includes multiple iterations of preparing, evolving, and measuring quantum states for optimizing a cost function, amplifies the effect of quantum noise. The accumulation of errors on each iteration hinders the ability to get accurate results.
As quantum computers get more complicated or require more qubits, these flaws accumulate, which further restricts their practical applicability. Accurately modeling and characterizing noise sources in quantum hardware is essential for understanding their effects and developing effective mitigation strategies.
While the success of SHARC-VQE under ideal conditions demonstrates its proof of principle, its real test lies in its performance under noisy conditions.

\begin{figure*}[!]
 \includegraphics[width=\textwidth]{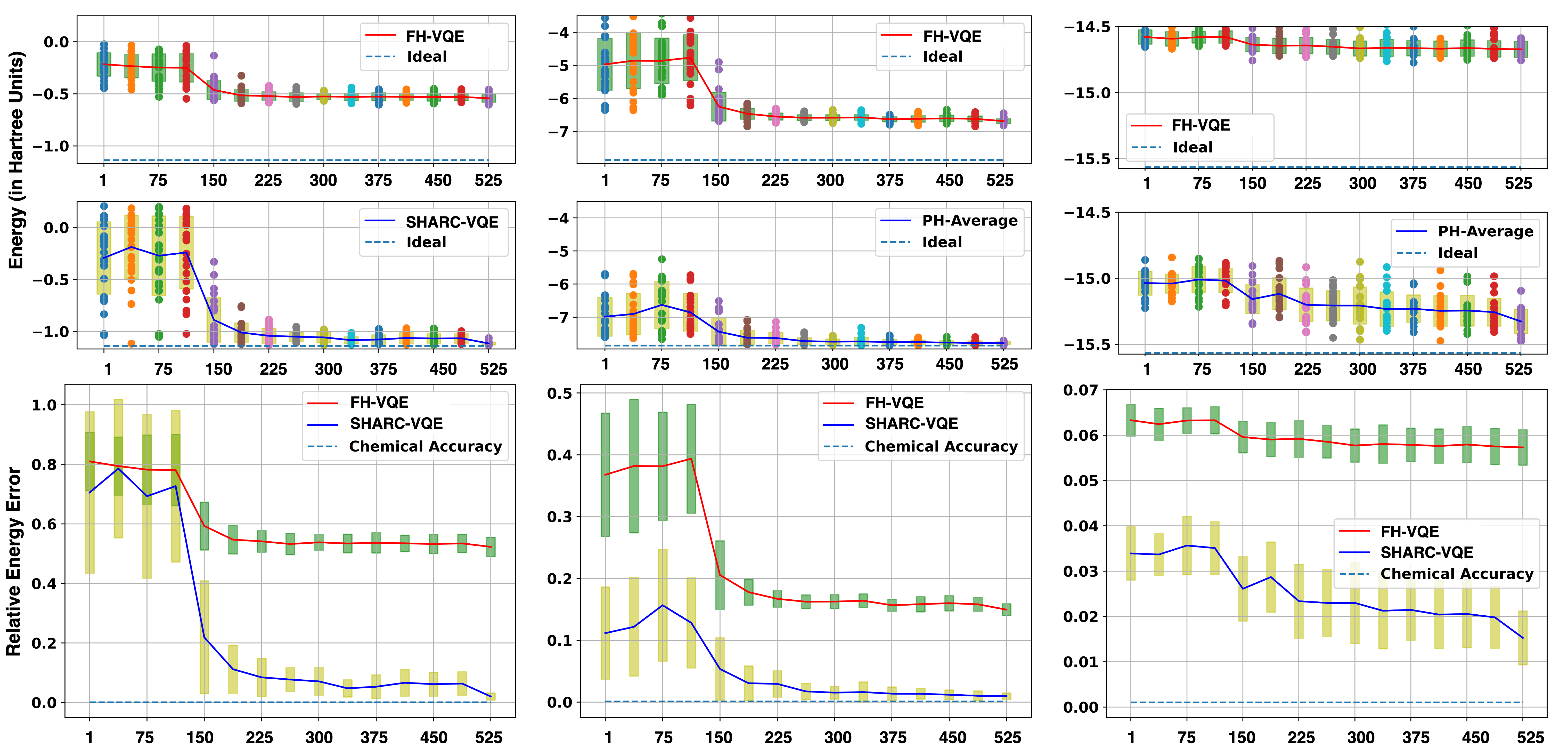} \\ 
 \includegraphics[width=\textwidth]{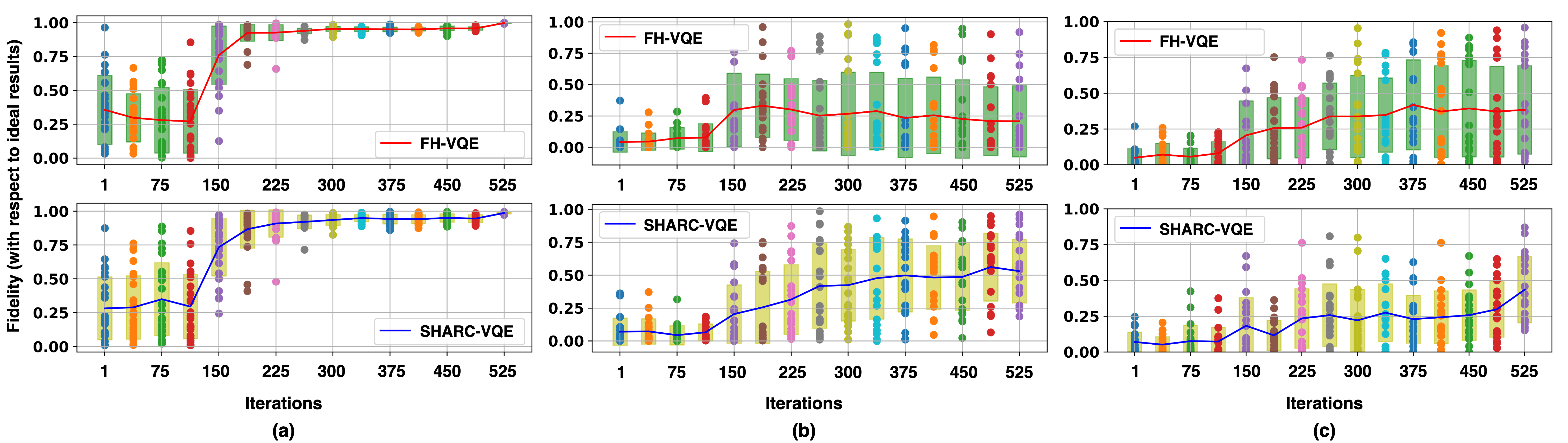}
 \caption{\label{fig:noiseenergy}The ground state energy (upper row), the relative energy with respect to the corresponding converged ground state energy of the full Hamiltonian (middle row) and the fidelity of the wave function with respect to the wave function of the full Hamiltonian (bottom row) for (a) H$_2$, (b) LiH, (c) BeH$_2$, and (d) H$_2$O under noisy conditions. The results shown are from 25 different VQE runs. In the case of the ground state energy and fidelity plots, the results of 25 VQE runs are shown as solid circles, with the average shown as a solid line and the standard deviation as a box.  Only the average and the standard deviation values are shown for the relative energy error.
}
\end{figure*}


Noise can be particularly damaging to chemistry applications with variational quantum algorithms. For a $N$-qubit Hamiltonian, the number of expectation values required to be evaluated scales as $\mathcal{O}(N^4)$. 
This massively increases error accumulation over a large number of measurements and consequent VQE iterations. This is where SHARC-VQE can be really helpful. Since all the Pauli strings in the partial Hamiltonian can be grouped, a single circuit measurement would suffice for energy evaluation. Hence, for a particular iteration of the simulation, the SHARC-VQE method effectively reduces the complexity from $\mathcal{O}(\frac{N^4}{\epsilon^2})$ to $\mathcal{O}(\frac{1}{\epsilon^2})$ for a certain desired precision $\epsilon$. 

FIG.~\ref{fig:noiseenergy} compares the performance of all three variants of SHARC-VQE under noisy conditions. For each system, 25 VQE simulations were run to sample the effect of noise on simulations adequately. The individual results, the average, and the standard deviations are highlighted in the figure. The SHARC-VQE technique outperforms the full Hamiltonian method, with the errors in energy dropping from $30-40\%$ (in the latter) to $5-10\%$ (in the former), see FIG.~\ref{fig:noiseenergy}. A similar performance is also seen in the fidelity calculations, where the fidelity is evaluated against the exact wavefunction obtained from the ideal quantum circuit evaluations, FH-VQE. The fidelity calculations, in particular, highlight the strength of the SHARC-VQE technique. SHARC-VQE not only estimates the energy more accurately, but it is also able to reproduce the wave function reliably. This shall have cascading benefits when molecular properties are evaluated from the quantum states of the system. It is noteworthy, that the excellent performance of the SHARC-VQE method is obtained without using any error mitigation technique, therefore, with no overhead cost.

\subsection{Hamiltonian Simplification Prior to Qubit Transformation}
In the present work, we have employed a Hamiltonian simplification approach after transforming the fermionic Hamiltonian (Equation~\ref{eq:8}) to its qubit form. However, the Hamiltonian simplification is also possible prior to the transformation. In the second quantized form of the Hamiltonian (Equation~\ref{eq:8}), the one- and two-electron integrals ($h_{pq}$ and $h_{pqrs}$ for H$_2$ given in 
Table~S2 in the supporting information) correspond to the one-electron operators $a^{\dagger}_ia_j$, and the two-electron operators $a^{\dagger}_ia^{\dagger}_ja_ka_l$, respectively. The qubit transformation (denoted by $\cal T$) of the one-electron operators to the corresponding Pauli strings, i.e., ${\cal T}(a^{\dagger}_i a_j) = \sum_{i} P_i$, results in Pauli strings with only I and Z Pauli gates when $i=j$. Hence, they can be categorized as easy-to-execute. Similarly, for the two-electron operator transformation (${\cal T}(a^{\dagger}_ia^{\dagger}_j a_k a_l) = \sum_{i} P_i$), when $(k,l) = (i,j) \times (i,j)$, the resulting qubit strings shall fall in the easy-to-execute category. 
Further, any operator with a repeated index (on either the creation or the annihilation operator, e.g., $a^{\dagger}_ia^{\dagger}_j a_k a_k$, or $a^{\dagger}_ia^{\dagger}_i a_k a_l$) shall vanish.
Any second-quantized molecular Hamiltonian can be simplified based on these criteria and then subjected to qubit transformation followed by evaluation using the SHARC-VQE approach. 
However, while constructing the partial Hamiltonian before qubit transformation,
the ease of computation can not be decided a priori. Here, one may ignore Pauli strings that are easy to execute, thus losing accuracy with no cost benefit. To counter this, the post-qubit transformation of the molecular Hamiltonian needs to be assessed to keep track of the gates being excluded in the partial Hamiltonian construction.

\subsection{Barren Plateau Problem in SHARC-VQE}
The barren plateau problem presents a significant challenge in the execution of variational quantum algorithms~\cite{bpintro1, bpintro2,McClean2018}. The barren plateau problem refers to a situation where the cost landscape becomes extremely flat, which poses a challenge for optimization algorithms in locating the global minimum. 
While the barren plateau problem is mainly linked to the size of the system, that is, the number of qubits, it can be linked to other hyperparameters of the quantum circuit, such as, the entanglement of the wave-function~\cite{bpentang} or its non-locality~\cite{bpwave} and the quantum noise~\cite{bpnoise}, making the current NISQ hardware highly susceptible. Addressing the barren plateau problem is crucial for effective application and scaling up of VQAs in addressing real-life issues~\cite{bpsol1, bpsol2, bpsol3, bpsol4, bpsol5}.

\begin{figure*}[!]
 \includegraphics[width=\textwidth]{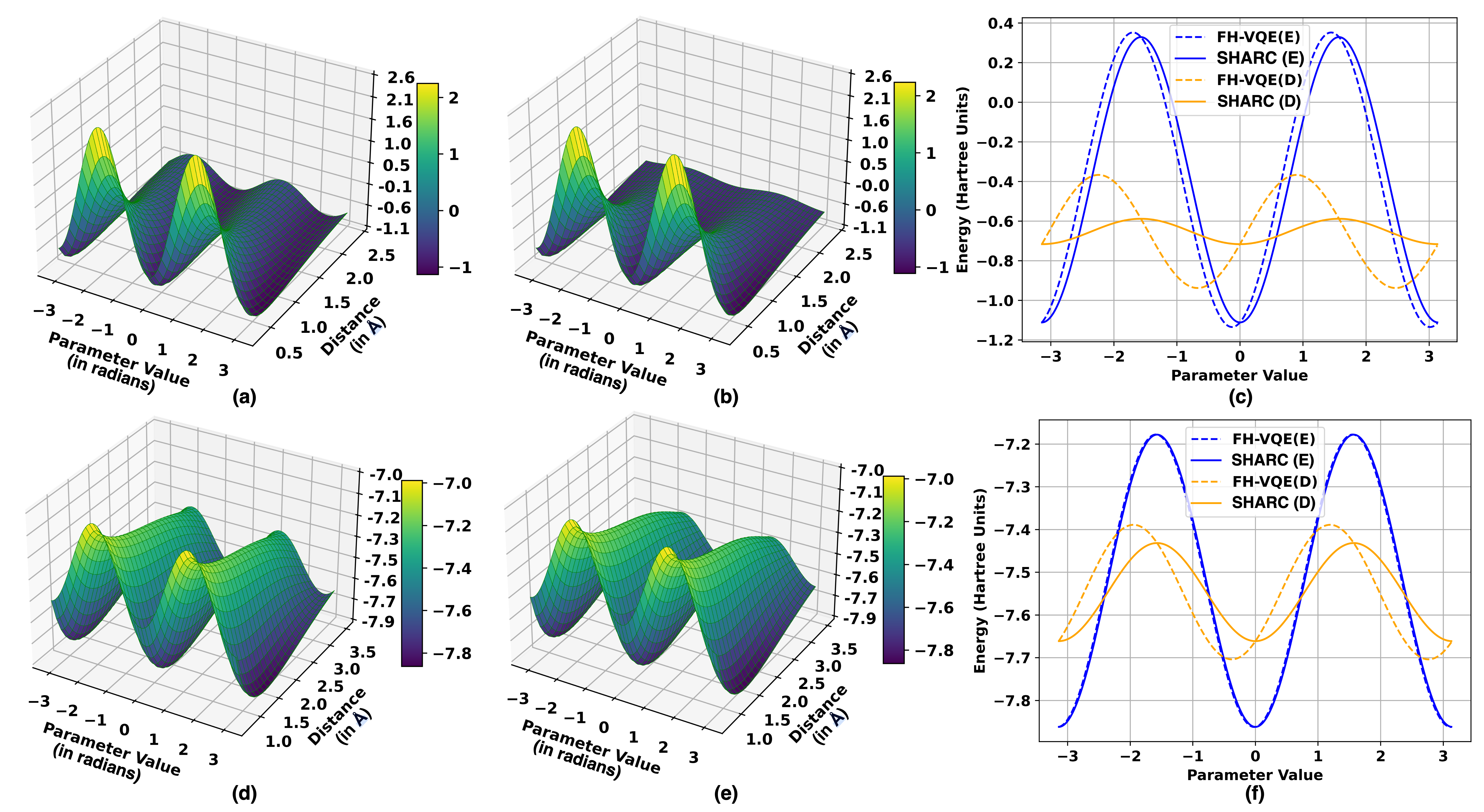}
 \caption{\label{fig:surface} The potential energy surface for H$_2$ ((a) full Hamiltonian and (b) SHARC-VQE) and LiH ((d) full Hamiltonian and (e) SHARC-VQE). A cross-section of the potential energy surface at the equilibrium geometry (at the internuclear distance of 0.74~\AA\ for H$_2$ and 1.59~\AA\ for LiH, referred to as E in the figure) and at the dissociated geometry (at the internuclear distance of 2.5~\AA\ for H$_2$ and 3.5~\AA\ for LiH referred to as D in the figure) for (c) H$_2$ and (f) LiH.}
\end{figure*}

While replacing a full Hamiltonian with a partial Hamiltonian simplifies the problem and intuitively makes it more efficient, a recent study finds that the simplified Hamiltonians are more likely to be susceptible to barren plateaus problem due to their complicated cost landscape~\cite{bpmain}. 
In that context, we explore the energy surfaces of different molecules and identify the consequences of the approximations made under SHARC-VQE.  
To illustrate this, we evaluate the potential energy surface (PES) of H$_2$ and LiH, each as a 4-qubit system initiated from their corresponding Hartree-Fock state. In these cases, the one-electron excitations do not contribute to the energy due to the Brillouin theorem~\cite{szabo}. Therefore, only one parameter corresponding to the single two-electron excitation is required. 
FIG.~\ref{fig:surface} shows the PES of H$_2$ and LiH as a function of the internuclear distances and the excitation parameter. The PES for both molecules were obtained using full Hamiltonian VQE and SHARC-VQE.  
While the overall cost landscape for the full and partial Hamiltonians appear similar (FIG.~\ref{fig:surface}), a closer comparison of the surface at the equilibrium and dissociated geometry reveals some interesting results (FIG.~\ref{fig:surface} c and f). 
At the equilibrium geometry, the potential-energy curves are nearly identical, barring the small shift in the minima in H$_2$, which gets smaller in LiH. However, SHARC-VQE yields considerably different results from the full Hamiltonian at the dissociated geometry. This could be because the approximation operator within the SHARC-VQE scheme was parameterized in the vicinity of the Hartree-Fock state at the equilibrium geometry, and the same parameterized operator was used at all other geometries, including the dissociated geometry. This assumption fails since the electronic structure at the dissociated geometry differs substantially from that at the equilibrium geometry. 

\subsection{SHARC-VQE as an Initialization Technique: The Fermi-Hubbard Model}\label{sec:initialization}

The Fermi-Hubbard model describes the dynamics of interacting fermions on a lattice structure. It serves as a fundamental framework for understanding the behavior of electrons in strongly correlated materials. In the Fermi-Hubbard Hamiltonian~\cite{fermihubbardmodel},
\begin{equation}
    H = \sum_{i,j} \sum_{\sigma = \uparrow, \downarrow} t_{ij} a^{\dagger}_{i\sigma} a_{j\sigma} + U\sum_{i} n_{i\uparrow}n_{i\downarrow}
\end{equation}
$t_{ij}$ represents the kinetic energy term associated with the hopping of an electron (in $\sigma$ spin-state) from a site $i$ to an adjacent site $j$, and $U$ accounts for the strength of onsite interactions among the electrons. The number operator $n_{i\sigma}$, defined as $a^{\dagger}_{i \sigma} a_{i\sigma}$, accounts for the occupancy of electron in $\sigma$ spin-state at the lattice site $i$. In this case, the chosen Fermi-Hubbard Hamiltonian was constructed with open boundary condition and a uniform interaction, $t_{ij} = -1.0$, while the onsite potential, $U$ was chosen to be 5.0. The resulting qubit Hamiltonian of the lattice is presented in TABLE~\ref{tab:fh}. Unlike the molecular Hamiltonians (Table~\ref{tab:h2full}), all terms in this Hamiltonian are significant. In this case, Hamiltonian simplification based on the significance of the terms is not straightforward. In such cases, a SHARC-VQE can be used as an initialization technique. One of the major challenges in the variational algorithms is to find a good starting point in the simulation~\cite{start1, start2}. The success of SHARC-VQE in reliably obtaining the full Hamiltonian's true ground state with high fidelity establishes it as an ideal initialization technique for a variety of other VQE problems. Initially, a simplified partial Hamiltonian (PH) is constructed only based on the ease of execution of the gates, and additional terms can be added as required. The extent of the excluded information from the partial Hamiltonian can be quantified as,
\begin{equation}
    k_{PH} = \frac{\sum^{m}_{i}|c_i|}{\sum^{n}_{j}|c_j|},
\end{equation}
where, the sum of the coefficients of the $m$ and $n$ Pauli strings in the partial and the full Hamiltonians, respectively, quantifies a degree of similarity between them. 
The optimized state from the partial Hamiltonian VQE is then used as the initial state for a new simulation, with the full Hamiltonian, in the so-called partial Hamiltonian initialized-VQE (PHI-VQE). 

\begin{table}[htbp]
\caption{\label{tab:fh} Hamiltonian of the Fermi Hubbard lattice in 4-qubit space with a general form (Equation~\ref{eq:sum_of_pdt_pauli}).}
\begin{center}
\begin{tabular}{ccc|ccc}
\hline
\hline
Index$^{\mathrm{a}}$  & Coefficient  & Gates & Index$^{\mathrm{a}}$  & Coefficient  & Gates\\
\hline
1    &  $+$2.5  & IIII  & 7     & $+$ 1.25 & ZZII\\
2    & $-$1.25   & IIZI  &  8   & $-$ 0.5 & IYZY\\
3    & $-$1.25   & IIIZ  &  9   & $-$ 0.5 & IXZX\\
4    & $+$ 1.25   & IIZZ  &  10   & $-$ 0.5 & YZYI\\
5    & $-$ 1.25   & ZIII  &  11   & $-$ 0.5 & XZXI\\
6    & $-$ 1.25   & IZII  &  \\
\hline
\multicolumn{6}{l}{$^{\mathrm{a}}$The sequence follows the Qiskit ordering.}
\end{tabular}
\end{center}
\end{table}

\begin{figure*}[!]
 \includegraphics[width=\textwidth]{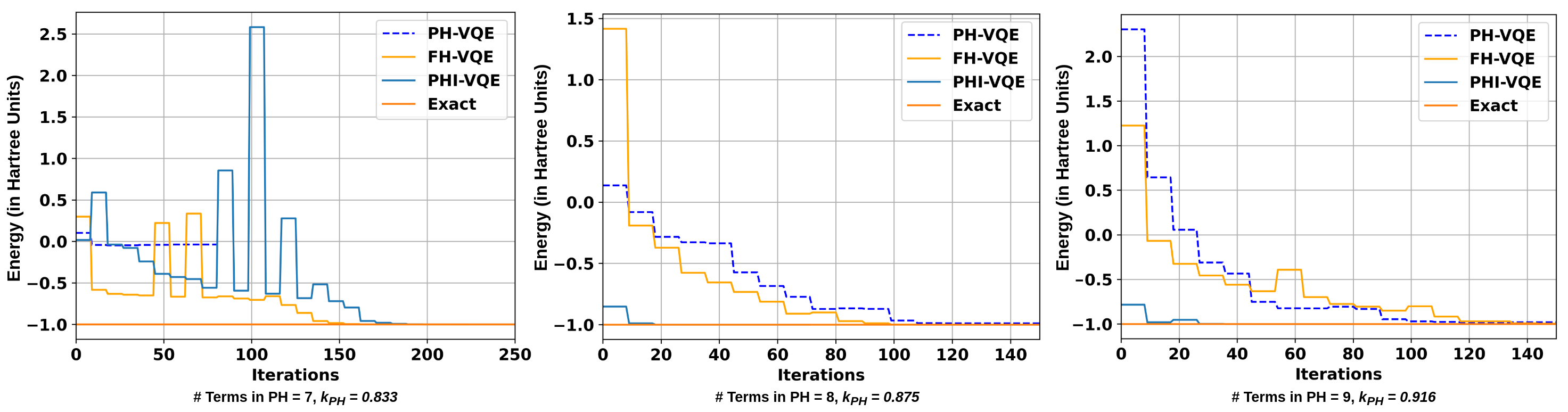}
 \caption{\label{fig:fermihubbard}Ground state energy under ideal quantum simulation settings for 2-node, 4-qubit Fermi-Hubbard lattice with varying number of terms in the partial Hamiltonian.  $k_{PH}=$ (a) 0.833, (b) 0.876, and (c) 0.916. The dashed lines highlight the result of the PH-VQE and the end of that simulation serves as the starting point for PHI-VQE.}
\end{figure*}

Figure~\ref{fig:fermihubbard} highlights the performance of the PHI-VQE method as an initialization technique for the Fermi-Hubbard model. The partial Hamiltonian prepared by just considering the easy-to-execute terms (7 out of 11 terms, with $k_{PH}=0.875$ fails to come close to the actual energy (Figure~\ref{fig:fermihubbard}a), whereas the addition of one or two more terms in the partial Hamiltonian improves the situation  (Figure~\ref{fig:fermihubbard}b and c). The energy approaches its true value in the latter cases, although the convergence rate is slower than the full Hamiltonian VQE. However, with a good starting point provided by the PH-VQE, the convergence is achieved in PHI-VQE within a few iterations (15-30), as opposed to the random initialized parameters ($>120$), see Figure~\ref{fig:fermihubbard}b and c. For the present model, we observe that once $k_{PH} \ge 0.85$, the PH-VQE method can reach a good initial state for the standard VQE. For different problems, the ideal value of $k_{PH}$ needs to be determined. However, more research on different problems is required for a general prescription of the optimized value of $k_{PH}$.

\subsection{Comparison of SHARC-VQE with other Methods}
There have been several efforts to enhance the performance of quantum computing for complex problems. In this section, we compare SHARC-VQE with other methods with shared objectives. McClean et al. have demonstrated that incorporating spatial locality with the Bravyi−Kitaev transformation leads to enhanced scalability of the existing quantum algorithms for quantum chemistry~\cite{McClean2016}. Considering the proximity of physical interactions, it is conceivable that numerous terms in the Hamiltonian can be deemed insignificant for a finite desired accuracy $\epsilon$. 
By exploiting the Gaussian product rule, i.e., the product of two spherical Gaussian functions is a rapidly decaying Gaussian function along the line segment connecting the two centers, one can reduce the overall scaling of the two-electron integrals to ${\cal O}(N^2)$. This approach, while laying the foundation for quadratic scaling methods of electronic structure, is basis-dependent. Another drawback of the locality-based method is that the insignificant two-electron integrals are not included in the VQE simulation, possibly making the method difficult to scale for larger molecules. Both of these drawbacks have been addressed in the SHARC-VQE method, which is basis-independent and introduces an approximation operator for the insignificant terms.

Basis rotation grouping~\cite{Huggins2021} relies on decomposing the two-body operator via tensor analysis. It provides a method to reduce the total number of terms, particularly the joint terms, to be measured in the Hamiltonian. This method, based on a two-stage decomposition of the interaction tensor~\cite{basrot2}, has been employed to minimize the overall gate depth of the whole UCCSD Ansatz, as well as the Trotter steps~\cite{basrot1}. This reduction results in a linear scaling with the size of the system, where the measurement is conducted on a distinct basis for each term. This requires an additional ${\cal O}(N)$ gate depth to execute the orbital rotation for each grouped term of the decomposed Hamiltonian before the measurement itself. 
The Hamiltonian partitioning scheme is based on the concept that a Pauli string's Abelian group can be diagonalized simultaneously by employing a unitary rotation of the measurement basis~\cite{group1, group2, group3, group4}. This process effectively reduces the number of terms to be measured to accurately calculate the expectation value of a Hamiltonian. Using the principles of the stabilizer theory, individual measurement values for all the terms in a particular Abelian group are evaluated by performing a single joint measurement on the whole qubit register. \par
Both the partitioning and grouping techniques help in reducing the overall number of Hamiltonian measurements but require additional quantum resources, generally of the order of ${\cal O}(N)$, hence reducing the overall complexity to ${\cal O}(N^2 \sim N^3)$. However, additional quantum gates lead to worse performance of the VQEs against noise. This was also addressed with the SHARC-VQE technique, where no additional quantum gates are required. \par
Other potential methods of reducing the measurement overhead of a molecular Hamiltonian include machine learning and tomography techniques. Intending to minimize the number of shots needed to attain a specific degree of accuracy in measuring a Hamiltonian, machine learning on a sequence of shot results has been used to lower the variability of the expectation value~\cite{Torlai2020}. A replica of the state is first generated by an ansatz by using an unsupervised restricted Boltzmann machine (RBM)~\cite{rbm1}. The replica is subsequently employed to calculate the expectation values of quantum observables. RBMs have demonstrated efficacy as effective models for representing quantum states in the domain of condensed matter physics~\cite{rbm2, rbm3}. By limiting the tomography problem~\cite{tomo1, tomo2} to the estimation of a Hamiltonian, the expense of measurements can be greatly reduced. Considering the Pauli weight of the Hamiltonian can aid in decreasing the sampling prerequisites. It is important to mention that the methods discussed here are primarily intended for Hamiltonian characterization, rather than ground state estimation. Therefore, they may not be immediately optimized for usage in the context of the VQE. This is why the exact advantage and computational complexity of these methods are not known. \par

Table~\ref{tab:compare} compares the strength of SHARC-VQE vis-a-vis other available methods of Hamiltonian approximation and measurement reduction. SHARC-VQE presents an easily applicable, and most efficient method to reduce the number of the Hamiltonian measurements without requiring any additional quantum resources. 

\begin{table*}
\caption{\label{tab:compare}Comparison of the divide and conquer method to other available methods of Hamiltonian approximation and measurement reduction}
\begin{ruledtabular}
\begin{tabular}{c|cccccc}

Technique & Requires & Extra Gates & Corrections   & Measurement  & Versatility & \#\\
 & Approximation & Required &  &  Reduction  &  & Measurements$^{\mathrm{a}}$\\

\hline  
\hline  

Locality-Based & \cmark & \xmark & \xmark & \cmark & \xmark & $\mathcal{O}(n^2)$ \\
Tomography-Based & \xmark & \cmark & - & \cmark & \xmark & - \\
Unitary-Partitioning & \xmark & \cmark & - & \cmark & \xmark & $\mathcal{O}(n^2 \sim n^3)$ \\
Group Rotations  & \cmark & \cmark & \xmark & \cmark & \xmark & $\mathcal{O}(n^2 \sim n^3)$ \\
 SHARC-VQE  & \cmark & \xmark & \cmark & \cmark & \cmark & $\mathcal{O}(1)$

\label{table4}
\end{tabular}
\end{ruledtabular}{$^{\mathrm{a}}$Per iteration of a VQE simulation. The overall complexity is in terms of the number of shots S, which scale as $(\frac{1}{\epsilon^2})$ for a particular precision $\epsilon$.}

\end{table*}

\subsection{Computational Advantage of SHARC-VQE}
Measurements of different Pauli strings, when not grouped together are independent and therefore uncorrelated, resulting in the mean squared error for the energy estimate for the qubit Hamiltonian (Equation~\ref{eq:sum_of_pdt_pauli})~\cite{mse1, mse2}, 
\begin{equation}
    \epsilon = \sqrt{\sum^{n}_{i} \frac{c^2_i \rm{Var}[\hat{P}_i]}{S_i}}
\end{equation}
where $S_i$ is the number of shots employed in the measurement of the Pauli string $P_i$. Since the Pauli matrices are self-inverse, the variance can be written as,
\begin{eqnarray}
\rm{Var}[\hat{P}_i] =&& \langle \psi | \hat{P}^2_i | \psi \rangle - \langle \psi | \hat{P}_i | \psi \rangle^2 \\
&& = 1 - \langle \psi | \hat{P}_i | \psi \rangle^2 \leq 1.
\label{eq:2}
\end{eqnarray}
Hence, the mean squared error, as a function of the number of shots becomes, 
\begin{equation}
    \epsilon = \sqrt{\sum^{n}_{i} c^2_i \bigg( \frac{1 - \langle \psi | \hat{P}_i | \psi \rangle^2}{S_i}\bigg) }.
\end{equation}

\begin{figure}[!]
 \includegraphics[width=0.5\textwidth]{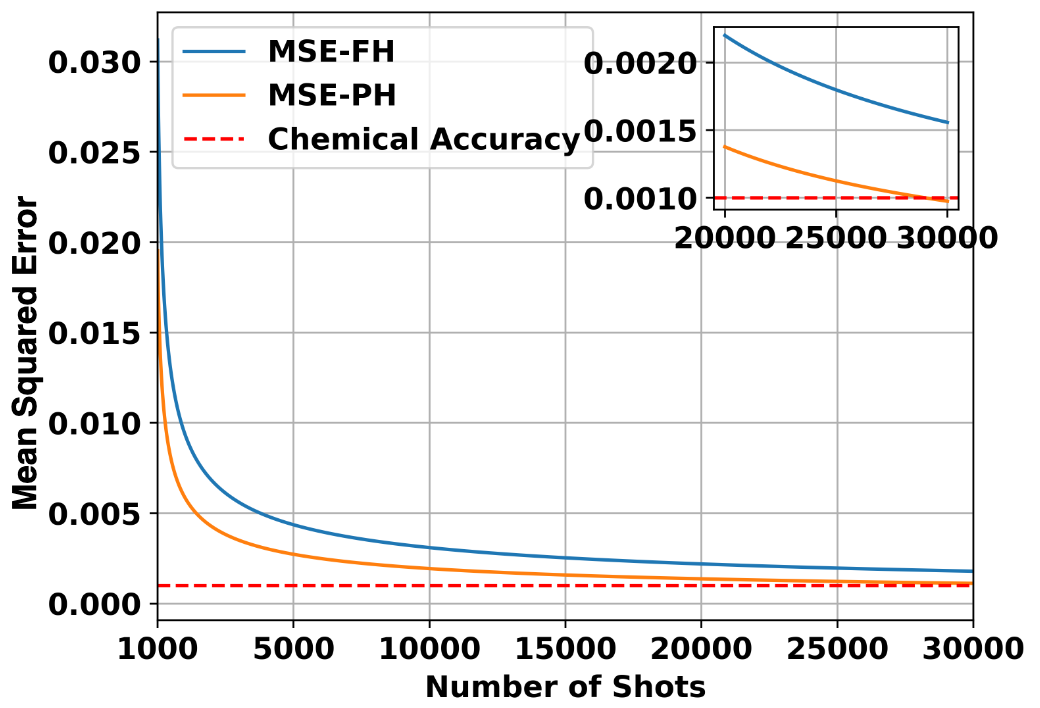}
 \caption{\label{fig:numPauli} Mean squared error in ground state energy calculation for the H$_2$ molecule as a function of the number of shots of the quantum circuit.}
\end{figure}

\begin{figure}[!]
 \includegraphics[width=0.5\textwidth]{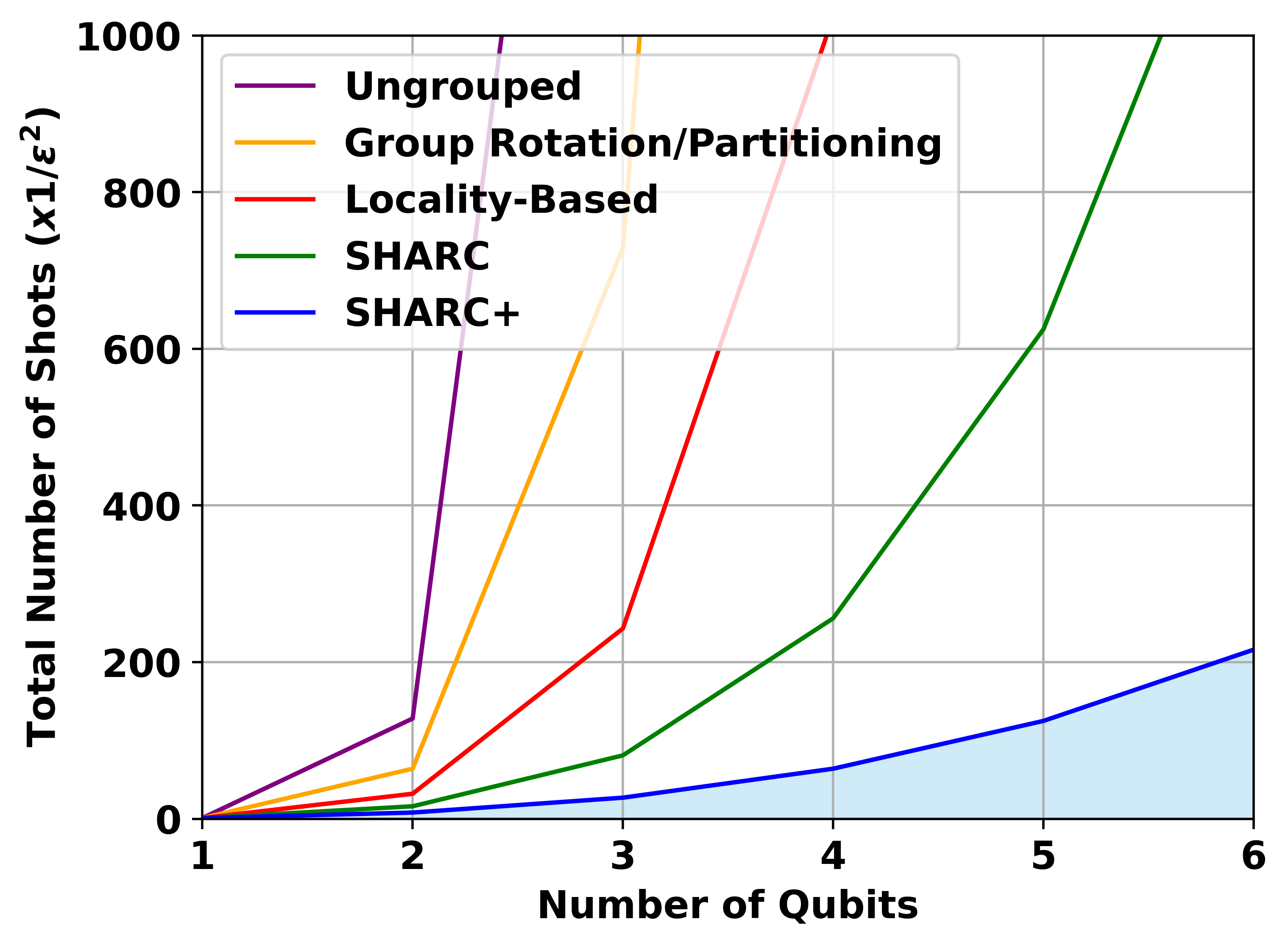}
 \caption{\label{fig:complex} Total number of shots for VQE simulation against the number of qubits in the system using different grouping techniques.}
\end{figure}

For the ungrouped Pauli strings, FIG.~\ref{fig:numPauli} shows the mean squared error in the energy calculation for H$_2$ against the number of shots of the quantum circuit, where SHARC-VQE shows a faster convergence. This performance would get better with larger systems. Interestingly, when grouping is considered, the computational advantage of SHARC-VQE is even better. Considering the fact that all the strings in the partial Hamiltonian contain either the `I', or, `Z' gate, they can be grouped together and a single circuit would be sufficient to calculate the energy. Hence, the number of Pauli string measurements can be reduced from $\mathcal{O}(\frac{n^4}{\epsilon^2})$ to $\mathcal{O}(\frac{1}{\epsilon^2})$ for a particular precision $\epsilon$ per VQE iteration. The other grouping or measurement reduction techniques can help reduce the complexity to $\mathcal{O}(\frac{n^3 \sim n^2}{\epsilon^2})$. Even then, SHARC-VQE remains computationally most inexpensive.

For a VQE simulation with a Hamiltonian having $n$ terms, a required precision of $\epsilon$ in $k$ number of optimizer iterations, the number of shots scale as,
\begin{equation}
    S \sim \mathcal{O}\bigg( \frac{kn}{\epsilon^2} \bigg).
\end{equation}
Scaling of the maximum number of optimizer iterations required with the number of terms in the Hamiltonian is not straightforward to evaluate. However, based on our previous study~\cite{hd}, the number of iterations needed scales worse than second-order with the number of qubits ($N$). In the best-case scenario, it can be assumed to be $k\sim \mathcal{O}(N^3)$. Hence, the overall computational complexity of the VQE problem as the function of the number of terms in the Hamiltonian can be written as,
\begin{equation}
     S \sim \mathcal{O}\bigg( \frac{N^3 n}{\epsilon^2} \bigg).
\end{equation}
For a general ungrouped Hamiltonian ($n \propto	N^4$), 
\begin{equation}
     S \sim \mathcal{O}\bigg( \frac{N^{7}}{\epsilon^2} \bigg)
\end{equation}
With other grouping techniques, $n$ can be reduced to  $n \propto  N^2 \sim N^3$. In that case, 
\begin{equation}
     S \sim \mathcal{O}\bigg( \frac{N^{5} \sim N^{6}}{\epsilon^2} \bigg).
\end{equation}
With SHARC-VQE, the $k$-iteration VQE simulation is done with the partial Hamiltonian, which can be measured with a single circuit. This makes $k$ scale linearly with the number of terms in the partial Hamiltonian $m \ (<n)$.
Since the partial Hamiltonian mainly consists of easy-to-execute terms, it nearly scales quadratically with the number of qubits. In SHARC-VQE, two additional measurements are required for the correction terms. Therefore we have,
\begin{equation}
     S_{\rm{SHARC}} \sim \mathcal{O}\bigg( \frac{N^3}{\epsilon^2} + \frac{2 N^4}{\epsilon^2} \bigg)
\end{equation}
It should be noted that other grouping techniques can still be employed for the corrections in the SHARC-VQE method (denoted by SHARC+). This further reduces the cost,
\begin{equation}
     S_{\rm{SHARC+}} \sim \mathcal{O}\bigg( \frac{N^3}{\epsilon^2} + \frac{2( N^2 \sim N^{3})}{\epsilon^2} \bigg).
\end{equation}
The superior scaling of SHARC-VQE methods is highlighted in Figure~\ref{fig:complex}. The reduced number of measurements helps in the error resistance of the SHARC-VQE algorithm, as highlighted by the improved performance against noise in Figure~\ref{fig:noiseenergy}. 

\section{Conclusions}
In this work, we have presented the SHARC-VQE (Simplified Hamiltonian approach with refinement and correction enabled variational quantum eigensolver) method, which offers a significant advancement in performing quantum chemical calculations on noisy intermediate-scale quantum devices. By partitioning the full molecular Hamiltonian into a more manageable Partial Hamiltonian and a less significant correction term, this method addresses the primary bottleneck of evaluating expectation values of a large number of Pauli strings in VQEs. The partial Hamiltonian approach introduced by SHARC-VQE reduces computational costs and enhances the accuracy and reliability of molecular simulations by mitigating noise from quantum circuits. SHARC-VQE performs exceptionally well in ideal quantum settings, converging to the ground state energy well within chemical accuracy with the full Hamiltonian VQE. SHARC-VQE also performs well in the presence of noise, outperforming the full Hamiltonian method, with the errors in energy dropping from $30-40\%$ (in the latter) to $5-10\%$ (in the former). SHARC-VQE is shown to reduce the computational costs for molecular simulations significantly. The cost of a single energy measurement can be reduced from $\mathcal{O}(\frac{N^4}{\epsilon^2})$ to $\mathcal{O}(\frac{1}{\epsilon^2})$ for a system of $N$ qubits and accuracy $\epsilon$. The overall cost of VQE can be reduced from $\mathcal{O}(\frac{N^7}{\epsilon^2})$ to $\mathcal{O}(\frac{N^3}{\epsilon^2})$, which is much better than the existing simplification techniques, scaling at best at $\mathcal{O}(\frac{N^5}{\epsilon^2})$. Furthermore, the method's utility as an initialization technique for the Fermi Hubbard model underscores its potential to optimize starting points for more elaborate quantum simulations. Overall, SHARC-VQE improves the efficiency, accuracy, and practicality of VQEs in quantum chemistry, paving the way for more effective and scalable quantum simulations of molecular properties.

\section*{Conflict of interest}
There are no conflicts to declare.

\section*{Acknowledgements}
This work used the Supercomputing facility of IIT Kharagpur established under the National Supercomputing Mission (NSM), Government of India, and supported by the Centre for Development of Advanced Computing (CDAC), Pune. HS acknowledges the Ministry of Education, Govt. of India, for the Prime Minister's Research Fellowship (PMRF).

\section*{Author Contribution} 
HS: Data Curation, Formal Analysis,  and  Original Draft Writing. S. Majumder and S. Mishra: Supervision, Review \& Editing.

\bibliography{main.bib}

\end{document}